\documentclass[a4paper,11pt]{article}
\usepackage{fancyheadings}
\usepackage[dvips]{graphicx}
\usepackage{ifthen}
\usepackage{amssymb}
\usepackage{amsmath}
\usepackage{epsfig}
\usepackage[errorshow]{tracefnt}
\newcommand\ffam{\sffamily}
\newcommand\fser{\bfseries}
\newcommand\fsh{\upshape}

\newcommand\sss{\scriptscriptstyle}
\newcommand\scs{\scriptstyle}

\newcommand\lp{\ell_p}

\newcommand\nn{\nonumber}
\newcommand\bea{\begin{eqnarray}}
\newcommand\eea{\end{eqnarray}}
\newcommand\Ht{\tilde{H}}
\newcommand\we{\wedge}

\newcommand\benu{\begin{enumerate}}
\newcommand\eenu{\end{enumerate}}
\newcommand\bit{\begin{itemize}}
\newcommand\eit{\end{itemize}}

\newcommand{\be}{\begin{eqnarray}}
\newcommand{\ee}{\end{eqnarray}}
\newcommand{\bd}{\begin{displaymath}}
\newcommand{\ed}{\end{displaymath}}
\newcommand{\bq}{\begin{equation}}
\newcommand{\eq}{\end{equation}}

\newcommand\RR{{\mathbb R}}

\newcommand\ZZ{{\mathbb Z}}

\newcommand\wdg{\wedge}

\newcommand\ga{\gamma}
\newcommand\de{\delta}

\newcommand\eps{\epsilon}

\newcommand\etal{{\it et al.} }
\newcommand\ie{{\it i.e.}}
\newcommand\eg{{\it e.g.}}
\usepackage{hyperref}
\begin{document}

%
%

\thispagestyle{empty}

\null\vskip-0.5cm

\begin{tabular}{l}
\epsfig{file=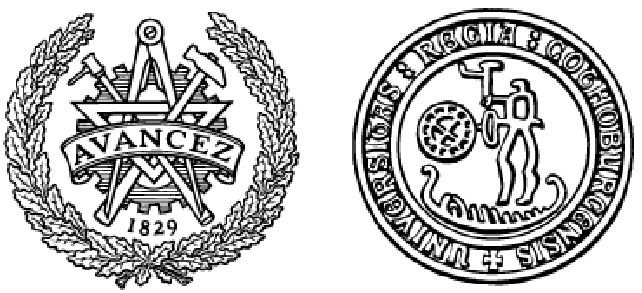,height=1.7cm}
\end{tabular}
\hfill
\begin{tabular}{r}
G\"oteborg ITP preprint\\
{\tt hep-th/0108113}\\
August 17, 2001
\end{tabular}\\
\vskip-2mm
\hrulefill
\vskip2cm

\begin{center}
{\upshape\sffamily\bfseries\huge On the Equivalence of Bound State Solutions} \\[4mm]
\end{center}


\vspace*{4mm}
\begin{center}
    {\ffam\fsh\Large Ulf Gran\footnote{E-mail: gran@fy.chalmers.se} and Mikkel Nielsen\footnote{E-mail: mikkel@fy.chalmers.se}}\\
\end{center}


\vspace{1cm}
\begin{center}
    {\ffam\fsh Department of Theoretical Physics\\*[1mm]
    G\"oteborg University and Chalmers University of Technology\\*[1mm]SE-412 96 G\"oteborg, Sweden}
\end{center}
\normalsize

%
%

\vspace{1cm}
\centerline{\ffam\fser Abstract}
\vspace{2mm}
In this paper we show the equivalence of various (non-threshold) bound state solutions of branes, or equivalently branes in background potentials,  in ten- and eleven-dimensional supergravity. We compare solutions obtained in two very different ways. One method uses a zero mode analysis to make an Ansatz which makes it possible to solve the full non-linear supergravity equations. The other method utilises T-duality techniques to turn on the fields on the brane. To be specific, in eleven dimensions we show the equivalence for the (M2,M5) bound state, or equivalently an M5-brane in a $C_{(3)}$ field, where we also consider the (MW,M2,M$2'$,M5) solution, which can be obtained from the (M2,M5) bound state by a boost. In ten dimensions we show the equivalence for the ((F,D1),D3) bound states as well as the bound states of $(p,q)$ 5-branes with lower dimensional branes in type IIB, corresponding to D3-branes in $B_{(2)}$ and $C_{(2)}$ fields and $(p,q)$ 5-branes in $B_{(2)}$, $C_{(2)}$ and $C_{(4)}$ fields. We also comment on the recently proposed V-duality related to infinitesimally boosted solutions.

%
%
\newpage
\section{Introduction}
A lot of papers have appeared on the subject of bound states of
branes\footnote{We will only consider the non-marginal half-supersymmetric bound states, where the lower dimensional branes lie within the higher dimensional brane.}, and the supergravity solutions can be obtained in several different but equivalent ways. The purpose of this paper is to establish this equivalence explicitly.

The most commonly used method utilises T-duality
techniques to turn on the fields on the brane, but alternatively
the solutions can be obtained by performing a zero mode analysis,
which enables one to make an Ansatz for the fields and then solve
the supergravity equations. This was done by Cederwall \etal
\cite{Adawi:1998,Cederwall:1998,Cederwall:1999}. Furthermore, the
solutions can be viewed either as a brane in background potentials 
or as a bound state of a brane and lower dimensional branes.
For instance, an (F,D$p$) bound state corresponds to a D$p$-brane
in an electric (\ie, a component including the time direction)
$B$-field background, and a (D($p-2$),D$p$) bound state corresponds to
a D$p$-brane in a magnetic $B$-field background. 

While the method of analysing the zero-modes generally gives a unique solution  parametrised by the field strength on the brane, the method of using T-duality techniques generally gives a multitude of solutions for the same brane. We argue that the method of analysing the
zero-modes always gives the most general half supersymmetric solution by
construction, since there is a unique way of fitting the half supersymmetric zero-modes into the
target space fields and since the subsequent full solution is uniquely obtained
from the zero-mode solution. Therefore, all the solutions obtained using T-duality techniques 
are related to solutions obtained by analysing the zero-modes and thus do not
generate a larger family of solutions despite the multitude of different looking
solutions. As a consequence of this, the most
general bound states in type IIB  was first
obtained in \cite{Cederwall:1998} for the D3-brane and in \cite{Cederwall:1999} for the $(p,q)$ 5-branes. 

The zero-mode method does not generate any waves in the solutions, but solutions
including waves \cite{Russo:1996,Bergshoeff:2000:3} have been shown to be related to solutions without waves
via finite boosts \cite{Russo:1996,Cai:2001}. Solutions with light-like fields have been shown to
be obtainable by taking a limit involving an infinite boost \cite{Alishahiha:2000,Cai:2000:2}.

The solutions have been seen to be important as the supergravity duals \cite{Hashimoto:1999,Maldacena:1999,Gopakumar:2000:1,Harmark:2000:1}  of noncommutative theories on branes \cite{Seiberg:1999,Gopakumar:2000:1,Gopakumar:2000:2,Bergshoeff:2000,Bergshoeff:2000:2} and since some papers on this subject use the solutions of Cederwall \etal as supergravity duals \cite{Berman:2000:1,Gran:2001:2} it also seems important to get an understanding of the relation between the different formulations.

The following is a chronological description of some of the important
papers\footnote{This is not intended to be a complete list.}. 
To our knowledge, the first explicit ten or eleven-dimensional supergravity solution, involving a tensor field on the brane, was obtained by Izquierdo \etal in 1995 \cite{Izquierdo:1995}. They construct an
M5-brane in a background of a 3-form potential, by lifting a known  eight
dimensional solution to eleven dimensions. The solution
corresponds to a bound state of M5- and M2-branes. In
1996 Russo and Tseytlin constructed the same solution, but for the
first time the solution was obtained using the mentioned duality
techniques \cite{Russo:1996}. In this paper, they also construct (F,D3), (D1,D3) and (D0,D2) bound states for the first time. Almost at the same time Breckenridge
\etal used the same method on D-branes, yielding (D($p-2$),D$p$)
bound states \cite{Breckenridge:1996}. Shortly afterwards Costa and Papadopoulos constructed the (F,D6) bound state (and they discuss the other (F,D$p$) bound states) \cite{Costa:1996}. In late 98, Cederwall \etal
constructed the most general SL(2,$\ZZ$)-covariant ((F,D1),D3) bound state for the first
time, using a zero mode analysis \cite{Cederwall:1998}, and the
(M2,M5) solution was also constructed in this way. The following
year Lu and Roy constructed (F,D$p$) bound
states (as well as the SL(2,$\ZZ$)-transformed states in type IIB
supergravity), using duality techniques
\cite{Lu:1999:1,Lu:1999:2,Lu:1999:3}. They also discuss the
(F,D($p-2$),D$p$) bound states. Concerning the 5-branes in type
IIB and the 2-form on them, they write down complete solutions (\ie,
including all the supergravity fields) in the cases where there is
an electric rank 2 field on the brane (in the rank 4 case, they only give the solution for the metric). The general rank D5- and NS5-brane
solutions were first obtained in the fall of 99 \cite{Alishahiha:1999}, but these
solutions were not complete (the D5 solution lacked the RR fields
and the NS5 solution lacked the NS-NS 2-form as well as the RR
4-form). The solutions correspond to the (F,D3,D5), (D3,D3,D5), (F,D3,D3,D5) bound states (and similarly for the NS5-brane). The solutions for general D$p$-branes with general rank of the $B$-field as well as the general NS5 solution in type IIA (corresponding to (D0,D2,NS5) and (D2,D4,NS5)) were also obtained (but again not with all the fields). In late 99 the most general $(p,q)$ 5-brane solution was constructed in
\cite{Cederwall:1999}. This includes the following bound states, (F,D5), (D3,D5), (F,D1,D3,D5), (D1,D3,D3,D5), (F,D1,D3,D3,D5) as well as the
SL(2,$\ZZ$) transformations of these (and the cases mentioned above are obtained as special cases). These states correspond to
the following $B$-field configurations: electric rank 2, magnetic
rank 2, electric and magnetic rank 4, magnetic rank 4 and rank 6.
The complete NS5-branes solutions in type IIA appeared in
\cite{Mitra:2000}. 

In section 2, we describe the Goldstone mechanism, which yields an understanding of the zero modes on a brane and enables us to make an Ansatz for the fields and solve the full non-linear supergravity equations. This is done for the M5-brane and after that the D3- and $(p,q)$ 5-branes are discussed. In section 3, we show the equivalence between various M5-brane solutions as well as for the D3- and $(p,q)$ 5-branes. We end with a conclusion in section 4.

\section{Zero modes}

\subsection{The M5-brane}

It is well known that massless degrees of freedom, so called Goldstone modes,
arise when a continuous symmetry is broken. Put, \eg, an M2-brane into an eleven
dimensional space. This breaks half of the supersymmetry, another half is broken
by the Dirac equation when going on-shell, resulting in eight
fermionic zero-modes. The translational symmetry in the transverse
directions is also broken, generating eight bosonic zero-modes. Since we get the same number
of bosonic and fermionic degrees of freedom, there is a supersymmetric theory
living on the M2-brane. The M2-brane case, however, contains only scalar modes,
which have been understood for quite some time. The situation is a bit different
if we instead look at the M5-brane. Here we get eight fermionic degrees of
freedom as above, but now we only get five bosonic degrees of freedom from the
breaking of translational symmetries. The three
extra bosonic degrees of freedom needed to get a supersymmetric theory come from
an anti-self-dual three-form field strength on the brane.   
In \cite{Adawi:1998}, the Goldstone mechanism is generalised to tensor fields of arbitrary
rank, providing the same level of understanding in terms of broken symmetries as for the scalar modes. Here we will just sketch the ideas.

The generalisation to tensor modes can be made by understanding how the scalar modes arise. Since we are
studying a theory with gravity, \ie, a theory having local diffeomorphism invariance, we
have to be more careful when we say that introducing a brane breaks
translational symmetry in the transverse directions. We can always make a {\em
small} diffeomorphism, \ie, a diffeomorphism taking the same value in the two
asymptotic regions of the brane solution, $r\rightarrow 0$ and $r\rightarrow
\infty$, without changing any conserved quantities like, \eg, the momentum. If we
instead make a {\em large} diffeomorphism, taking different values in the
asymptotic regions, we change conserved quantities and it is therefore these
symmetries that are broken in the presence of a brane. Since diffeomorphisms are
the gauge symmetry associated with gravity, we come to the conclusion that
Goldstone modes are associated with broken large gauge symmetries. In the case
of the fermionic modes it is the large supersymmetry transformations that are
broken and, \eg, the tensor modes on the M5-brane come from broken large gauge
transformations of the background three-form potential. 

 By doing a ``rigid'' transformation, \ie, an $x$-independent gauge transformation, where $x$ denotes the longitudinal coordinates, we get information on
how to introduce the zero-modes in the relevant field. By {\em then} turning on the
$x$-dependence, to obtain a theory on the brane, we can get the equations of motion for the zero-modes by using the supergravity field equation, since after turning on the $x$-dependence we are no longer considering just a gauge transformation. To illustrate this method we will now analyse the tensor modes on the M5-brane \cite{Adawi:1998}. Using the notation of \cite{Adawi:1998}, we make
a gauge transformation of the background three-form potential $\de C=d\Lambda$,
where $\Lambda=\Delta^k A$ and $A$ is a {\em constant} two-form which lies in
the transverse directions since we want a
theory on the brane and $k$ is a constant which will be determined from the equations of motion. We first calculate $\de
C=d\Delta^k\we A$ and {\it then} turn on the $x$-dependence of
$A$, after which the variation of $C$ is no longer just a gauge transformation and we can therefore obtain equations of motion for the zero-modes by using the supergravity equations. We can now compute the variation of the four-form field strength $H=dC$,
\begin{equation}
h=\de H=d\Delta^k\we F
\end{equation}
where $F=dA$. The field equation for $H$ is, to linear order in $h$,
\begin{equation}
d\ast h-H\we h=0
\end{equation}
By inserting the M5-brane solution for $H$ we get
\begin{equation}
\Delta d\ast_x F\we \ast_y d\Delta-(k\ast_x F-F)\we d\Delta\we \ast_y d\Delta=0
\end{equation}
where $\ast_x$ and $\ast_y$ denote dualisation in the longitudinal and
transverse directions respectively. By considering the two duality components of
$F$ separately (fulfilling $\ast_x F=\pm F$) we get that $k=-1$ for the
anti-self-dual part and $k=1$ for the self-dual part. We also get the equation
of motion $d\ast_x F=0$. Since each duality component of $F$ contributes with three
bosonic degrees of freedom, we seem to have twice the number of extra degrees of
freedom that we needed in order to get supersymmetry. However, since we want a theory on the brane, we must require that the zero modes are normalisable when integrating out the transverse directions.
By doing this, we see that the self-dual part of $F$ has non-normalisable zero-modes,
and must therefore be discarded. We have thus seen how the tensor modes on the
M5-brane can be understood as arising from broken large gauge transformations of
the background three-form potential.

\subsection{Type IIB branes}
Just as for the M5-brane, we can do a zero mode analysis for the 3- and 5-branes in type IIB \cite{Adawi:1998,Cederwall:1999}, but here we get complications due to the SL(2,$\ZZ$) symmetry. As discussed in detail in \cite{Adawi:1998}, the additional bosonic zero modes on D-branes, correspond to vector modes and the deformed supergravity solutions are then parametrised by the corresponding 2-form field strength on the brane. These zero modes arise when we break the large gauge transformations of the background 2-form potentials. In a manifestly SL(2,$\ZZ$)-covariant formulation, we have a doublet of 2-forms which can be combined into a complex 2-form, see appendix A for details.
To be specific, the deformations are parametrised by a complex anti-selfdual 2-form $F_{(2)}$ on the D3-brane \cite{Adawi:1998} and by a real 2-form $F$ on the $(p,q)$ 5-branes \cite{Cederwall:1999}. This also yields a matching of the number of fermionic and bosonic zero modes. In both cases we have 8 fermionic zero modes. On the D3-brane we have 6 scalar zero modes and the last two exactly correspond to half of the modes for a complex vector in 4 dimensions (and the half is due to the anti-selfduality of the field strength). On the 5-branes, we have 4 scalars, and the remaining 4 bosonic zero modes correspond to the number of degrees of freedom for a real vector in 6 dimensions.

In general, we can start from any brane, whose normalisable zero-modes we identify using the
prescription described above. This gives us exact knowledge of how the zero-modes appear in
all target space fields and enables us to make an Ansatz for the full
solution. The non-linear supergravity equations are then solved for the unknown
functions in the Ansatz. The results are presented in the following sections. Apart from the general M5-brane solution, only
special cases of these solutions were known prior to \cite{Cederwall:1998,Cederwall:1999}.

\section{Equivalence of solutions}\label{relations}

\subsection{The M5-brane}\label{m5}
In this section we show the equivalence of various M5-brane solutions. We will start from the solution obtained in \cite{Cederwall:1998}, representing an (M2,M5)
bound state, and derive an explicit mapping to the (M2,M5) solution of Izquierdo
\etal \cite{Izquierdo:1995}, who were the first to obtain the (M2,M5) solution. 
We will then boost the
solution of Izquierdo \etal leading to the (MW,M2,M$2'$,M5) solution of Bergshoeff
\etal \cite{Bergshoeff:2000:3}, essentially following \cite{Cai:2001} but using a slightly
generalised form of the mapping.

The (M2,M5) solution obtained in \cite{Cederwall:1998} is\footnote{Note that we have chosen a
sign such that $h_{+++}=-h_{---}=-\sqrt{2\nu}$ in the formulation of \cite{Cederwall:1998}.}
\begin{equation}
\begin{split}
ds^2&=(\Delta_+\Delta_-)^{1/3}\left[\Delta_-^{-1}\left(-dt^2+(dx^1)^2+(dx^2)^2\right)\right.\\
&\qquad\left.+\Delta_+^{-1}\left((dx^3)^2+(dx^4)^2+(dx^5)^2 \right)+dr^2+r^2d\Omega^2_4\right]\\
\lp^3 C_3&=\sqrt{2\nu}\left[\Delta_-^{-1}dt\wedge dx^1\wedge
dx^2-\Delta_+^{-1}dx^3\wedge dx^4\wedge dx^5\right]\\
H_4&=dC_3+3\pi N \epsilon_4
\end{split}
\end{equation}
where $\lp$ is the eleven dimensional Planck length, $N$ is the number of
M5-branes in the bound state, $\epsilon_4$ is the volume element on the unit
4-sphere and
\begin{equation}
\Delta=k+\left(\frac{R}{r}\right)^3\,,\qquad R\equiv \pi N^{1/3}\lp
\end{equation}
is the harmonic function and $\Delta_\pm=\Delta\pm\nu$, 
where $\nu$ is proportional to the square of the field strength on the brane, see
\cite{Cederwall:1998} for details. We
must have $\nu\leq k$ in order to avoid naked singularities.
The critical case $\nu=k$, discussed at the end of this section, is related to the supergravity dual of OM theory.
The (M2,M5) solution of Izquierdo \etal \cite{Izquierdo:1995} is 
\begin{eqnarray}
ds^2&=&H^{-1/3}h^{-1/3}\left[-dt^2+(dx^1)^2+(dx^2)^2+h\left((dx^3)^2+(dx^4)^2+(dx^5)^2
   \right)\right. \nonumber \\
&&\left.+H(dr^2+r^2 d\Omega^2_4)   \right]\nonumber\\
\lp^3 C&=&H^{-1}\sin \alpha\, dt\wedge dx^1\wedge dx^2-H^{-1}h \tan \alpha\,
dx^3\wedge dx^4 \wedge dx^5\label{Izquierdo}\\
\nonumber
H_4&=&dC_3+ 3\pi N \epsilon_4\nonumber
\end{eqnarray}
where the function $h$ and the harmonic function $H$ are defined as
\begin{equation}
H=A+\frac{R^3}{\cos\,\alpha\, r^3}\,,\qquad h^{-1}=H^{-1}\sin^2\alpha+\cos^2\alpha
\end{equation}
where we have allowed for an arbitrary constant $A$ in the harmonic function.

The equivalence can be seen by making the following substitutions
\begin{align}\label{rel}
\begin{split}
\frac{\Delta_-}{2\nu}&=\frac{H}{\tan^2\alpha}\\
\frac{\Delta_+}{2\nu}&=\frac{H}{h \sin^2\alpha}\\
\frac{(k-\nu)}{2\nu}&=\frac{A}{\tan^2\alpha}
\end{split}
\end{align}
keeping $R$ unchanged and rescaling the coordinates according to
\begin{align}
\begin{split}
r&\rightarrow\left(\frac{\tan^2\alpha\cos\alpha}{2\nu}\right)^{1/3}r\\
x^{0,1,2}&\rightarrow\left(\frac{2\nu\cos^2\alpha}{\tan^2\alpha}\right)^{1/6}x^{0,1,2}\label{scaling}\\
x^{3,4,5}&\rightarrow\left(\frac{2\nu}{\tan^2\alpha\cos^4\alpha}\right)^{1/6}x^{3,4,5}
\end{split}
\end{align}
where $x^0\equiv t$. Note that we can not solve for $A$ and $\alpha$ in terms of $k$ and $\nu$, we just have the relation in (\ref{rel}). Usually one chooses $A=k=1$ and then the relation determines $\alpha$ in terms of $\nu$. This, however, prevents us from being able to map the critical solutions in a non-singular manner, which will be important later on.

Having shown the equivalence of the (M2,M5) solution, we now turn to the boosted solutions.  We perform the boost
\begin{equation}
t\rightarrow t\cosh\gamma-x^5\sinh\gamma\,,\qquad x^5\rightarrow x^5\cosh\gamma-t\sinh\gamma\label{lorentz}
\end{equation}
on the solution of Izquierdo \etal (\ref{Izquierdo}), for details see
\cite{Cai:2001}. Instead of the parameters $\alpha$ and $\gamma$, we can use the angles $\theta_1$ and $\theta_2$, with $\theta_1\leq \theta_2$, defined by 
\begin{equation}
\cos\alpha=\frac{\cos\theta_2}{\cos\theta_1}\,,\;\;\;\cosh\gamma=\frac{\sin\theta_2}{\cos\theta_1}\frac{1}{\sin\alpha}\,,\;\;\;\sinh\gamma=\frac{\sin\theta_1}{\cos\theta_2}\frac{1}{\tan\alpha}\label{angles}
\end{equation}
The boosted solution becomes
\begin{equation}
\begin{split}
ds^2&=(H' h_1 h_2)^{-1/3}\bigg[-dt^2+h_1\left((dx^1)^2+(dx^2)^2\right)+h_2\left
((dx^3)^2+(dx^4)^2\right)\\
&\qquad+h_1 h_2\left(dx^5+\sin\theta_1\sin\theta_2({H'}^{-1}-1)dt\right)^2+H'(dr^2+r^2d\Omega_4^2)\bigg]\\
\lp^3 C_3&={H'}^{-1}\left(\frac{\sin\theta_2}{\cos\theta_1}h_1 dt\we dx^1\we
dx^2+\frac{\sin\theta_1}{\cos\theta_2}h_2 dt\we dx^3\we dx^4 \right.\\        
&\qquad-h_1\tan\theta_1dx^1\we dx^2\we dx^5-h_2\tan\theta_2 dx^3\we dx^4\we dx^5\bigg)\\
H_4&=dC_3+3\pi N\epsilon_4
\end{split}
\end{equation}
with the harmonic functions
\begin{equation}
H'=B+\frac{R^3}{\cos\theta_1 \cos\theta_2\,
r^3}\,,\qquad h_i^{-1}={H'}^{-1}\sin^2\theta_i+\cos^2 \theta_i
\end{equation}
which satisfy
\begin{equation}
H=H' h_1^{-1}\,,\qquad h^{-1}=h_1 h_2^{-1}
\end{equation}
and we have allowed for an arbitrary constant in the harmonic function which is related to $A$ as follows
\begin{equation}
B=\frac{A-\sin^2\theta_1}{\cos^2\theta_1}
\end{equation}
Note that for $A=1$ we get $B=1$, which is the case considered in \cite{Cai:2001}.

We now want to match this to the solution of Bergshoeff \etal \cite{Bergshoeff:2000:3}
\bea
ds^2&=&(E_1 E_2)^{1/3}\left[-\Ht^{-1}\left[1-(1-\Ht)^2\frac{s_1^2 s_2^2}{E_1 E_2}\right]dt^2\right.\nn\\
&&+\frac{2}{E_1 E_2}(1-\Ht)s_1 s_2 dt dx^5+\frac{\Ht}{E_1 E_2}(dx^5)^2+\frac{1}{E_1}\left((dx^1)^2+(dx^2)^2\right)\nn\\
&&\left.+\frac{1}{E_2}\left((dx^3)^2+(dx^4)^2\right)+dr^2+r^2d\Omega_4^2\right]\label{BCOT}\\
dC_3&=&d\left(\frac{1-\Ht}{E_1}\right)c_1 s_2\we dt\we dx^1 \we
dx^2+d\left(\frac{1-\Ht}{E_2}\right)c_2 s_1\we dt\we dx^3\we dx^4\nn\\
&&-d\left(\frac{1-\Ht}{E_1}\right)c_1 s_1\we dx^1\we dx^2\we
dx^5-d\left(\frac{1-\Ht}{E_2}\right)c_2 s_2\we dx^3\we dx^4\we dx^5
\nn\\
&&-c_1 c_2\star d\Ht\nn
\eea
where $s_i=\sin\theta_i$, $c_i=\cos\theta_i$ and 
\begin{equation}
E_i=s_i^2+\Ht c_i^2
\end{equation}
with the harmonic function
\begin{equation}
\Ht=a+\left(\frac{\tilde{R}}{r}\right)^3
\end{equation}
We have allowed for a constant $\tilde{R}$ in order not to have to rescale
the radial coordinate in the mapping.
The mapping is obtained by setting
\begin{equation}
H'=\Ht\,,\qquad H' h_i^{-1}=E_i
\end{equation}
without any coordinate rescalings.
The first requirement implies that
\begin{equation}
\tilde{R}^3=\frac{R^3}{\cos\theta_1\cos\theta_2}
\end{equation}
and also that the constants in the harmonic functions must be related, \ie,
\begin{equation}
a=\frac{A-\sin^2\theta_1}{\cos^2\theta_1}
\end{equation}
As in \cite{Cai:2001}, we have obtained a complete mapping between the boosted solution of Izquierdo \etal and the solution of Bergshoeff \etal \cite{Bergshoeff:2000:3} but now with arbitrary constants in the harmonic functions. 

We end this section with a discussion of the supergravity dual of OM theory where light membranes are obtained from a critical 3-form, here corresponding to $k=\nu$. Recently, possible decoupled theories of the boosted solution were discussed in \cite{Cai:2001}. The coordinate rescalings between the unboosted solutions can be written as
\bea
x^{0,1,2}&\rightarrow&\left(\frac{(k-\nu)}{A}\left[2\nu\left(\frac{A}{k-\nu}\right)+1\right]^{-1}\right)^{1/6}x^{0,1,2}\nn\\
x^{3,4,5}&\rightarrow&\left(\frac{(k-\nu)}{A}\left[2\nu\left(\frac{A}{k-\nu}\right)+1\right]^{2}\right)^{1/6}x^{3,4,5}\\
r&\rightarrow&\left(\frac{A}{(k-\nu)}\left[2\nu\left(\frac{A}{k-\nu}\right)+1\right]^{-1/2}\right)^{1/3}r\nn\\
\eea
If we want to be able to relate critical solutions we must require that the above mapping is non-singular when $\nu\rightarrow k$. We therefore let $A\rightarrow 0$ as
$\nu\rightarrow k$ in such a way that the quotient $A/(k-\nu)$ is kept fixed. Note also that the critical case $A=0$ corresponds to 
\begin{equation}
a=-\tan^2\theta_1\label{critcase}
\end{equation}
and therefore, as soon as we consider a boosted solution (see (\ref{angles})), the value of $a$ which
yields the critical solution is negative. Taking into account the change in $a$
there is no problem associated with boosting a critical solution and in particular a Lorentz transformation does not change the critical or non-critical aspect of a solution, as expected. 
For the above value of $a$, the asymptotic metric will be that of a smeared membrane,
just as in the usual supergravity dual of OM-theory. We have seen that
it is crucial to have arbitrary constants in the harmonic functions in order to be able
to have a non-singular mapping, thereby being able to relate critical solutions, in particular when the solutions are boosted.

We end this section with some comments on the recently proposed V-duality \cite{Chen:2000,Cai:2000:3,Cai:2001}. In \cite{Cai:2001} it is argued that it is
only possible to obtain a decoupled OM theory from an infinitesimally boosted
(M2,M5) bound state, but not from a finitely boosted one.
Important for this conclusion is that the decoupling limit is assumed to be 
the same after the boost as before.
We see that the restriction to infinitesimal boosts, \ie, galilean
transformations, follows directly from this assumption regarding the decoupling
limit. The decoupling limit is obtained by scaling $t\sim\eps^0$ and $x^5\sim\eps^{3/2}$
and demanding this scaling both before and after the boost gives, when inserted
into the Lorentz transformation (\ref{lorentz}), that $\sinh \ga\sim\eps^{3/2}$, \ie, $\ga\sim\eps^{3/2}$.
The restriction to infinitesimal Lorentz transformations can therefore be seen
to arise due to the different scalings of the coordinates when one tries to keep
the decoupling limit fixed, which amounts to an extra constraint on the system. 
If we instead use the decoupling limit from 
\cite{Berman:2000:1}, where all the coordinates on the M5-brane scale in the same
way, we do not get any restriction to infinitesimal boosts and V-duality reduces to ordinary Lorentz transformations.

\subsection{The D3-brane}\label{D3}

In this section, we present the translation between two D3-brane solutions, the one by Cederwall \etal \cite{Cederwall:1998} and the one by Lu and Roy \cite{Lu:1999:2}\footnote{We will use the version given in \cite{Lu:2000}.}. The first of these is an SL(2,$\RR$)-covariant description of all the bound states of a D3-brane and (F,D1)-strings, whereas the second is written as a specific but general ((F,D1),D3) bound state. When comparing the solutions, we can pick a certain bound state, and the general equivalence follows from SL(2,$\RR$)-covariance.

As mentioned in the previous section, the solution of \cite{Cederwall:1998} is described by a complex anti-selfdual 2-form $F_{(2)}$. The radial dependence of the undeformed solution is described by the harmonic function
\begin{equation}
\Delta=1+\frac{R^4}{r^4}
\end{equation}
We can then define the deformed harmonic functions $\Delta_\pm=\Delta\pm\nu$, where $\nu$ describes the deformation. More precisely, $\nu=2\vert\mu\vert$, where $\mu={\scs \frac{1}{4}}\mbox{{\rm tr}}\,\big(F_{(2)}\big)^2$. Then the solution is
\begin{eqnarray}\nonumber
ds^2&=&\Delta_+^{\frac{1}{4}}\Delta_-^{-\frac{3}{4}}\Big(-(dx^0)^2+(dx^1)^2\Big)+\Delta_+^{-\frac{3}{4}}\Delta_-^{\frac{1}{4}}\Big((dx^2)^2+(dx^3)^2\Big)+\Delta_+^{\frac{1}{4}}\Delta_-^{\frac{1}{4}}dy^2\\
{\cal H}_{(3)}&=&\big(\Delta_+\Delta_-)^{-\frac{1}{4}}d\Delta\wdg\Big(\Delta_-^{-\frac{3}{2}}\Pi_+ F_{(2)}+\Delta_+^{-\frac{3}{2}}\Pi_- F_{(2)}\Big)\\\nonumber
P&=&i\,\mu\,\big(\Delta_+\Delta_-\big)^{-1}d\Delta\,,\quad\quad Q=0
\end{eqnarray}
where $y^m\,,m=1,2,\ldots,6$ are the coordinates transverse to the D3-brane, $P$ and $Q$ are the Mauer-Cartan 1-forms, see appendix A for details, and the following projectors have been defined
\begin{equation}
(\Pi_\pm)_{\mu\nu}={\scriptstyle \frac{1}{2}}\Big(1\pm{\scriptstyle{\frac{2}{\nu}}}F_{(2)}\bar{F}_{(2)}\Big)_{\mu\nu}
\end{equation}
The solution is an SL(2,$\RR$)-covariant description of a bound state of a D3-brane and (F,D1)-strings. The charges of the strings are specified by the doublet of 3-forms. Choosing to get a specific bound state therefore corresponds to choosing the scalar doublet (from the set of possible solutions). Since $F_{(2)}$ is antisymmetric and anti-selfdual, we can use a basis in which it takes the following form
\begin{equation}
F_{(2)}=\sqrt{\mu}\left(\begin{array}{cccc}0&1&0&0\\-1&0&0&0\\
0&0&0&-i\\0&0&i&0\end{array}\right)
\end{equation}
We can then find a certain scalar doublet, corresponding to a bound state of an F-string and a D3-brane, or equivalently a solution with an electric NS-NS 2-form. From the doublet of scalars we get the dilaton and the axion as well as the doublet of 3-forms which can be integrated to yield the doublet of 2-forms
\begin{eqnarray}\nonumber\label{d3}
{\cal U}^1&=&-{\scs \frac{1}{c}}\,\eta\,\Delta_+^{-\frac{1}{4}}\Delta_-^{\frac{1}{4}}\,,\qquad\qquad\ \
{\cal U}^2=i\,c\,\eta\,\Delta_+^{\frac{1}{4}}\Delta_-^{-\frac{1}{4}}\\
C_{(2)1}&=&c\sqrt{2\nu}\,\Delta_-^{-1}\,dx^0\wdg dx^1\,,\quad
C_{(2)2}=c^{-1}\sqrt{2\nu}\,\Delta_+^{-1}\,dx^2\wdg dx^3\\\nonumber
e^\phi&=&c^{2}\sqrt{{\scs \frac{\Delta_+}{\Delta_-}}}\,,\qquad\qquad\qquad\quad\ \ \chi=0
\end{eqnarray}
where $c$ is an arbitrary real constant (actually $c^2$ is the undeformed asymptotic dilaton) and $\eta=\mu/\vert\mu\vert$. This solution has string charges ($p^1,p^2$)=($p$,0). Arbitrary string charges and background scalars can be obtained by performing an SL(2,$\RR$)-transformation
\begin{eqnarray}
\left(\begin{array}{c}p\\q\end{array}\right)=\left(\begin{array}{cc}1&p\tilde{p}\\q/p&p\tilde{q}\end{array}\right)\left(\begin{array}{c}p\\0\end{array}\right)\end{eqnarray}
where $\tilde{p},\tilde{q}$ are real numbers fulfilling $p\tilde{q}-q\tilde{p}=1$. Through such a transformation we can thus, \eg, get a non-vanishing axion. 

It is the (F,D3) bound state above we will use for comparison with the solution of Lu and Roy. Omitting
the transverse 5-form, which is not important in this context, their
solution is\footnote{There is a small typo in the 
solution in \cite{Lu:2000}, a missing factor of $g_s$ in $A_2$, as can be seen when
checking the SL(2,$\RR$)-covariance of the solution.}
\begin{eqnarray}\nonumber\label{luroy}
&&ds^2=g_s^{-1/2}\Big(H^{-\frac{3}{4}}H^{'\frac{1}{4}}\big(-d\tilde{x}_0^2+d\tilde{x}_1^2\big)+H^{\frac{1}{4}}H^{'-\frac{3}{4}}\big(d\tilde{x}_2^2+d\tilde{x}_3^2\big)+(HH')^{\frac{1}{4}}d\tilde{y}^2\Big)\nonumber\\
&&e^\phi=g_s\frac{H''}{\sqrt{HH'}},\quad\chi=\frac{pq(H-H')+g_s\chi_0\Delta_{(p,q)}H'}{q^2H+g_s^2(p-\chi_0
q)H'}\\\nonumber 
&&2\pi\alpha'B=g_s^{1/2}(p-\chi_0
q)\Delta_{(p,q,n)}^{-1/2}H^{-1}d\tilde{x}^0\wedge d\tilde{x}^1-\frac{q}{n}\,H^{'-1}d\tilde{x}^2\wedge d\tilde{x}^3\\\nonumber
&&A_2=g_s^{-1/2}\big(qg_s^{-1}-\chi_0(p-\chi_0
q)g_s\big)\Delta_{(p,q,n)}^{-1/2}H^{-1}d\tilde{x}^0\wedge d\tilde{x}^1+\frac{p}{n}\,H^{'-1}d\tilde{x}^2\wedge d\tilde{x}^3\\\nonumber
\end{eqnarray}
where
\begin{equation}
H=1+\frac{Q_3}{\tilde{r}^4} ,\quad
H'=1+\frac{n^2g_s^{-1}}{\Delta_{(p,q,n)}}\frac{Q_3}{\tilde{r}^4} ,\quad
H''=1+\frac{g_s^{-1}(q^2+n^2)}{\Delta_{(p,q,n)}}\frac{Q_3}{\tilde{r}^4}
\end{equation}
and
\begin{eqnarray} &\Delta_{(p,q)}=g_s(p-\chi_0 q)^2+g_s^{-1}q^2
,\quad\Delta_{(p,q,n)}=\Delta_{(p,q)}+g_s^{-1}n^2&\\\nonumber
&Q_3=4\pi\alpha^{'2}\Delta_{(p,q,n)}^{1/2}g_s^{3/2}&
\end{eqnarray}
Here $p$ and $q$ are the charges of the strings and $n$ is the number of
D3-branes lying on top of each other. The solution can also be written in terms of two angles
\begin{eqnarray}
&\cos\theta=\frac{n}{\sqrt{q^2+n^2}} ,\quad \cos\alpha=\frac{\sqrt{q^2+n^2}}{\sqrt{g_s^2(p-\chi_0 q)^2+q^2+n^2}}&
\end{eqnarray}
The cases $p-\chi_0 q=0$ ($q=0$) correspond to $\alpha=0$ ($\theta=0$) respectively, and these special cases are of the usual form used when constructing the solutions via T-duality and boosts/rotations.
To see the equivalence
between the two solutions, we only need to compare the (F,D3)
bound states with vanishing axion, since the general solution is
just obtained by an SL(2,$\RR$)-transformation. The form of the metric is the same independently
of the charges, so this part of the translation can be done for
the general configuration. From the metrics, the harmonic
functions can be identified \begin{eqnarray}\label{deltah}
&&\Delta_{-}=(1-\nu)\big(1+{\scs \frac{R^4}{(1-\nu)\,r^4}}\big)=(1-\nu)H\\\nonumber
&&\Delta_{+}=(1+\nu)\big(1+{\scs \frac{R^4}{(1+\nu)\,r^4}}\big)=(1+\nu)H'
\end{eqnarray}
We get exactly the same metric by doing the following coordinate transformation
\begin{eqnarray}\label{tilde}\nonumber
&&x^{0,1}=(1-\nu)^{\frac{3}{8}}(1+\nu)^{-\frac{1}{8}}g_s^{-\frac{1}{4}}\,\tilde{x}^{0,1}\\
&&x^{2,3}=(1-\nu)^{-\frac{1}{8}}(1+\nu)^{\frac{3}{8}}g_s^{-\frac{1}{4}}\,\tilde{x}^{2,3}\\\nonumber
&&r=(1-\nu)^{-\frac{1}{8}}(1+\nu)^{-\frac{1}{8}}g_s^{-\frac{1}{4}}\, \tilde{r}
\end{eqnarray}
And we also get a relation between the parameters governing the deformation
\begin{eqnarray}\label{alfanu}
&\cos^2\!\theta \cos^2\!\alpha=\frac{n^2}{g_s^2(p-\chi_0 q)^2+q^2+n^2}=\frac{1-\nu}{1+\nu}&
\end{eqnarray}
Using these relations in the special case of the (F,D3) bound state with vanishing axion, corresponding to $q=\chi_0=0$, we get exact agreement for all the fields, and as mentioned, the general case follows from the SL(2,$\RR$)-covariance. Thus it has been demonstrated that the two ((F,D1),D3) solutions indeed are equivalent.

A comment regarding the application of the above solution for supergravity duals
of noncommutative theories on branes: Using our solution, critical electric
field and infinite magnetic deformation parameter, $\tan\theta$, are obtained in
the limit $\nu\rightarrow 1$. The coordinate transformation (\ref{tilde})
therefore becomes singular in this limit. This is not a problem in itself, it
will just change the scaling of the coordinates with respect to
$\alpha'$. However, one has to be careful, if one wants to boost the critical
solutions. In particular, one has to keep track of which coordinates should be
fixed when $\alpha'$ goes to zero. If one wants to relate critical solutions, in particular when the solutions are boosted, the
coordinate transformation between the different formulations is required to be
non-singular in the critical limit $\nu=1$.
Hence (\ref{tilde}) has to be modified in this case, and the result will
be similar to that obtained in the previous section for the M5-branes, where we
had to allow for arbitrary integration constants in the harmonic functions in
order to be able to map between critical solutions in a non-singular manner. The same remarks of course hold for the $(p,q)$ 5-branes in the next section.

\subsection{$(p,q)$ 5-branes}\label{pq5}

In this section we will show the equivalence of various 5-brane
solutions in type IIB supergravity. In the rank six case, our
solution \cite{Cederwall:1999} is the only one containing $(p,q)$
5-branes; other papers only consider the D5 and/or the NS5
solutions. We will therefore compare our D5-brane solution with
that of \cite{Alishahiha:1999}. The latter solution does not
contain the RR fields, though.  To be able to compare all the
supergravity fields, we therefore also consider the rank two case
\cite{Cederwall:1999,Alishahiha:2000}. In the rank 2 case, general
$(p,q)$ 5-branes were first obtained in \cite{Lu:1999:3}, and
just as in the previous section, by relating two specific solutions the equivalence then follows from
SL(2,$\RR$)-covariance.

Our solution in \cite{Cederwall:1999} looks quite complicated, but it can actually be simplified a bit, by using the basis mentioned in section 5 of that paper. From the zero mode analysis we know that the solution is parametrised by a real 2-form $F$ which then takes the form
\begin{equation}
F=\left(\begin{array}{cccccc}
0&\tilde{\nu}_1&0&0&0&0\\
-\tilde{\nu}_1&0&0&0&0&0\\
0&0&0&\tilde{\nu}_2&0&0\\
0&0&-\tilde{\nu}_2&0&0&0\\
0&0&0&0&0&\tilde{\nu}_3\\
0&0&0&0&-\tilde{\nu}_3&0
\end{array}\right)
\end{equation} 
Define the harmonic functions 
\begin{equation}\label{defD}
\Delta_{\sss{\pm\pm\pm}}=\Delta\pm\nu_1\pm\nu_2\pm\nu_3
\end{equation}
where
$\nu_i$ corresponds to ${\scs \frac{9}{8}}\,\tilde{\nu}_i^2$ in
\cite{Cederwall:1999}. The solution contains expressions for the
complex 3-form as well as the 5-form field strengths. From the
scalar doublet we can get the doublet of 3-forms, and by integration
we obtain the 2- and 4-form potentials. The solution in the Einstein
frame takes the following form \begin{eqnarray}\label{5metric}\nonumber
&ds^2=\Delta_{\sss{---}}^{-3/4}(\Delta_{\sss{++-}}\Delta_{\sss{+-+}})^{1/4}
\big(-dx_0^2+dx_1^2\big)+\Delta_{\sss{++-}}^{-3/4}(\Delta_{\sss{---}}
\Delta_{\sss{+-+}})^{1/4}\big(dx_2^2+dx_3^2\big)&\\\nonumber
&\hspace{1cm}+\
\Delta_{\sss{+-+}}^{-3/4}(\Delta_{\sss{---}}\Delta_{\sss{++-}})^{1/4}
\big(dx_4^2+dx_5^2\big)+(\Delta_{\sss{---}}\Delta_{\sss{++-}}\Delta_{\sss{+-+}})^{1/4}dy^2&\\
&(C_{(2)r})_{01}=-2k p_r\sqrt{\nu_2\nu_3}\Delta_{\sss{---}}^{-1}-k^{-1}{\tilde p}_r\sqrt{2\nu_1}\,\Delta_{\sss{---}}^{-1}&\\\nonumber
&(C_{(2)r})_{23}=2k p_r\sqrt{\nu_1\nu_3}\Delta_{\sss{++-}}^{-1}-k^{-1}{\tilde p}_r\sqrt{2\nu_2}\,\Delta_{\sss{++-}}^{-1}&\\\nonumber
&(C_{(2)r})_{45}=2k p_r\sqrt{\nu_1\nu_2}\Delta_{\sss{+-+}}^{-1}-k^{-1}{\tilde p}_r\sqrt{2\nu_3}\,\Delta_{\sss{+-+}}^{-1}&
\end{eqnarray}
where $k$ is a real constant related to the asymptotic scalars, $(p_1,p_2)=(p,q)$ are the 5-brane charges and $\tilde{p}_r$ is a real doublet, fulfilling $\epsilon^{rs}p_r\tilde{p}_s=1$.
We write the 5-form as $H_5=d\Delta\wedge G_4$ plus the hodge dual and this 4-form
has the following components
\begin{eqnarray}
(G_{(4)})_{2345}=-\sqrt{2\nu_1}\,(\Delta_{\sss{++-}}
\Delta_{\sss{+-+}})^{-1}\nonumber\\
(G_{(4)})_{0145}=-\sqrt{2\nu_2}\,(\Delta_{\sss{---}}
\Delta_{\sss{+-+}})^{-1}\\
(G_{(4)})_{0123}=-\sqrt{2\nu_3}\,(\Delta_{\sss{---}}
\Delta_{\sss{++-}})^{-1}\nonumber
\end{eqnarray}
We can then integrate to get the 4-form potential. We have to make a
distinction between different ranks of the 2-form. For rank 6 we get
(the rank 4 case is just obtained by setting one of the $\nu$'s to zero)
\begin{eqnarray}\nonumber
(C_{(4)})_{2345}&=&{\scs \frac{\sqrt{2\nu_1}}{2(\nu_2+\nu_3)}}\,\log{\scs \frac{
\Delta_{+-+}}{\Delta_{++-}}}\\
(C_{(4)})_{0145}&=&{\scs \frac{\sqrt{2\nu_2}}{2(\nu_1+\nu_3)}}\,\log{\scs \frac{
\Delta_{---}}{\Delta_{+-+}}}\\\nonumber
(C_{(4)})_{0123}&=&{\scs \frac{\sqrt{2\nu_3}}{2(\nu_1+\nu_2)}}\,\log{\scs \frac{
\Delta_{---}}{\Delta_{++-}}}
\end{eqnarray}
In the rank 2 case we get by, \eg, putting $\nu_1$ and $\nu_2$ equal to zero,
\begin{eqnarray}\nonumber
(C_{(4)})_{2345}&=&(C_{(4)})_{0145}=0\\
(C_{(4)})_{0123}&=&\sqrt{2\nu_3}\,\Delta_{-}^{-1}
\end{eqnarray} 
The general expressions for the scalars can be found from the general scalar doublet in \cite{Cederwall:1999}\footnote{In that paper the solution is expressed in terms of three functions, $f_{{\sss 2-}}$, $f_{{\sss \mathrm{det}}}$ and $f_{{\sss 4}}$ which can be written like:  $f_{{\sss 2-}}=\Delta_{+--},\ f_{{\sss \mathrm{det}}}=\Delta_{+++}\Delta_{++-}\Delta_{+-+},\ f_{{\sss 4}}=\Delta_{+++}\Delta_{---}+4\nu_2\nu_3$.}. We will do the comparison
for $(p,q)=(0,1)$ (and therefore $\tilde{p}=-1$). In this special
case, the scalars take the following simple form \begin{equation}
e^{\phi}=k^{-2}\Big(\Delta_{---}\Delta_{++-}\Delta_{+-+}\Big)^{-\frac{1}{2}}\Delta_{+--}\,,\quad\chi=\tilde{q}-k^2\sqrt{8\nu_1\nu_2\nu_3}\Delta_{+--}^{-1}
\end{equation} So $k^{-2}$ is the undeformed asymptotic dilaton.

Now turn to the solution of \cite{Alishahiha:1999}, which can be obtained, using T-duality techniques. The solution is given in euclidean space. We can Wick rotate to obtain the lorentzian solution. The solution is furthermore given in the string frame. In the Einstein frame we get
\begin{eqnarray}\nonumber
ds^2=&&g_s^{-\frac{1}{2}}\Big(({\scs \frac{f}{h_1}})^{-\frac{3}{4}}({\scs \frac{f}{h_2}})^{\frac{1}{4}}({\scs \frac{f}{h_3}})^{\frac{1}{4}}\,d{\bf \tilde{x}}_{0,1}^2+({\scs \frac{f}{h_1}})^{\frac{1}{4}}({\scs \frac{f}{h_2}})^{-\frac{3}{4}}({\scs \frac{f}{h_3}})^{\frac{1}{4}}\,d{\bf \tilde{x}}_{2,3}^2\\&&+\quad({\scs \frac{f}{h_1}})^{\frac{1}{4}}({\scs \frac{f}{h_2}})^{\frac{1}{4}}({\scs \frac{f}{h_3}})^{-\frac{3}{4}}\,d{\bf \tilde{x}}_{4,5}^2+({\scs \frac{f}{h_1}}{\scs \frac{f}{h_2}}{\scs \frac{f}{h_3}})^{\frac{1}{4}}\,d\tilde{y}^2\Big)\\\nonumber
B_{01}=&&\tanh\theta_1 f^{-1}h_1\,,\quad B_{23}=\tan\theta_2 f^{-1}h_2\\\nonumber
B_{45}=&&\tan\theta_3 f^{-1}h_3\,,\qquad e^{2\phi}=g_s^2 f^{-1}h_1 h_2 h_3
\end{eqnarray}
where for simplicity, a shorthand notation is used for the line elements. Since the solution is lorentzian, a minus sign is understood in front of the time component in the metric.
Note that $g_s$ is the asymptotic value of the deformed dilaton.
The harmonic functions are given by
\begin{eqnarray}\nonumber
f&=&1+{\scs \frac{R^2}{\tilde{r}^2\,\cosh\theta_1\cos\theta_2\cos\theta_3}}\,,\quad h_1^{-1}=-\sinh^2\theta_1 f^{-1}+\cosh^2\theta_1\\
h_2^{-1}&=&\sin^2\!\theta_2 f^{-1}+\cos^2\!\theta_2\,,\qquad h_3^{-1}=\sin^2\!\theta_3 f^{-1}+\cos^2\!\theta_3
\end{eqnarray}
From the form of the metric and the dilaton we get the following relations between the
harmonic functions
\begin{eqnarray}\nonumber\label{harm}
&\Delta_{---}=(1\!-\!\nu_1\!-\!\nu_2\!-\!\nu_3)\,\frac{f}{h_1}\,,\quad \Delta_{++-}=(1\!+\!\nu_1\!+\!\nu_2\!-\!\nu_3)\,\frac{f}{h_2}&\\
&\Delta_{+-+}=(1\!+\!\nu_1\!-\!\nu_2\!+\!\nu_3)\,\frac{f}{h_3}\,,\quad \Delta_{+--}=(1\!+\!\nu_1\!-\!\nu_2\!-\!\nu_3)\,f&
\end{eqnarray}
We get exact agreement between the solution if we do the following coordinate transformation
\begin{eqnarray}\label{xtilde}\nonumber
&&x^{0,1}=g_s^{-1/4}(1\!-\!\nu_1\!-\!\nu_2\!-\!\nu_3)^{\frac{3}{8}}(1\!+\!\nu_1\!+\!\nu_2\!-\!\nu_3)^{-\frac{1}{8}}(1\!+\!\nu_1\!-\!\nu_2\!+\!\nu_3)^{-\frac{1}{8}}\,\tilde{x}^{0,1}\\\nonumber
&&x^{2,3}=g_s^{-1/4}(1\!-\!\nu_1\!-\!\nu_2\!-\!\nu_3)^{-\frac{1}{8}}(1\!+\!\nu_1\!+\!\nu_2\!-\!\nu_3)^{\frac{3}{8}}(1\!+\!\nu_1\!-\!\nu_2\!+\!\nu_3)^{-\frac{1}{8}}\,\tilde{x}^{2,3}\\\nonumber
&&x^{4,5}=g_s^{-1/4}(1\!-\!\nu_1\!-\!\nu_2\!-\!\nu_3)^{-\frac{1}{8}}(1\!+\!\nu_1\!+\!\nu_2\!-\!\nu_3)^{-\frac{1}{8}}(1\!+\!\nu_1\!-\!\nu_2\!+\!\nu_3)^{\frac{3}{8}}\,\tilde{x}^{4,5}\\
&&r=g_s^{-1/4}(1\!-\!\nu_1\!-\!\nu_2\!-\!\nu_3)^{-\frac{1}{8}}(1\!+\!\nu_1\!+\!\nu_2\!-\!\nu_3)^{-\frac{1}{8}}(1\!+\!\nu_1\!-\!\nu_2\!+\!\nu_3)^{-\frac{1}{8}}\, \tilde{r}
\end{eqnarray}
and we also get the relations between the parameters in the two formulations
\begin{eqnarray}\nonumber
&\cosh^2\!\theta_1=\frac{1+\nu_1-\nu_2-\nu_3}{1-\nu_1-\nu_2-\nu_3}\,,\quad \cos^2\!\theta_2=\frac{1+\nu_1-\nu_2-\nu_3}{1+\nu_1+\nu_2-\nu_3}&\\
&\cos^2\!\theta_3=\frac{1+\nu_1-\nu_2-\nu_3}{1+\nu_1-\nu_2+\nu_3}&
\end{eqnarray}

Now turn to the rank two case. In \cite{Alishahiha:2000}, the 4-form is also included in the NS5-brane solution with a rank 2 RR 2-form.
\begin{equation}
A_{0123}=g_s^{-1}\sin\theta\,f^{-1}
\end{equation}
The metric is the same as before with $\nu_1$=$\nu_2$=0 and therefore we have $\Delta_-=(1-\nu)f$. The translation of the metric, the NS-NS 2-form and the dilaton from above, of course also holds in this special case. Using the above relations between the coordinates and the parameters, restricted to the rank 2 case, we also get exact agreement for the 4-form.

Thus in the rank two case, it has been shown by comparing all the fields that our solution for the D5-brane is equivalent with the ones obtained by using T-duality techniques. As for the D3-brane, we also get a match in the general $(p,q)$ 5-brane case, since we just need to do an SL(2,$\RR$)-transformation.  In the rank six case, the equivalence has also been shown, but not all fields could be compared and not all 5-branes, since our solution is the only complete and SL(2,$\RR$)-covariant one in this case.

\section{Conclusion}
In this paper we have shown the equivalence of various bound
state solutions, obtained in different ways. Using our method, a
zero mode analysis is performed, enabling one to write down an
Ansatz and then solve the supergravity equations
exactly. With the other method, the fields on the brane are
obtained via T-duality techniques. The latter method is the
easiest one in the simplest cases, since one just has to use the T-duality rules
\cite{Buscher:1988,Buscher:1987,Bergshoeff:1995,Meessen:1998}. On
the other hand, we automatically get the most general solution by
finding the allowed zero modes on the brane. In their original
form, the solutions \cite{Cederwall:1998,Cederwall:1999} look
somewhat different from the ones obtained with T-duality
techniques, but by rewriting the solutions in a particular basis, the
resemblance between the solutions is increased. By performing a
rescaling of the coordinates and relating the harmonic functions
and the deformation parameters, we have shown that the different
methods indeed yield equivalent solutions.

For the solutions obtained through duality techniques, the $B$-field is obtained by T-dualising in a direction on the brane, rotating or boosting in the directions where one wants the $B$-field and then T-dualising back. Alternatively, one can do a double T-duality, a gauge transformation to get the $B$-field and then another double T-duality to obtain these solutions \cite{Hashimoto:1999,Berman:2000:2,Sundell:2000,Larsson:2001:2}. This method of course also yields equivalent solutions and the translation can be done in precisely the same way as in section 3.  

We have also clarified some issues regarding the (M2,M5)-brane and
its boosted solutions. In particular considering the decoupled theories, we have
shown that the V-duality requirement of infinitesimal boosts follows directly from
the assumption that the coordinates scale in the same way before and after the boost despite the fact that the boost mixes coordinates that scale differently.
Without any extra assumptions, \ie, using coordinates on the brane which scale homogeneously
in the limit giving the OM supergravity dual, as done in
\cite{Berman:2000:1}, we do not get any restrictions on the boost and V-duality reduces to ordinary Lorentz transformations. We have also
shown that there is no problem associated with boosting critical solutions when
taking into account the transformation of the constant in the harmonic function.   

\vspace{5mm}
{\Large {\bf Acknowledgements}}
\vspace{5mm}\\
We would like to thank M. Cederwall, B.E.W. Nilsson, H. Larsson and P. Sundell for valuable discussions. This work is partly supported by the Danish Natural Science Research Council.

\appendix

\section{Type IIB supergravity}

Type IIB supergravity in ten dimensions has an SL(2,$\RR$) invariance (which is broken to SL(2,$\ZZ$) by quantum effects) and contains the following fields: the metric, 2 scalars (the dilaton $\phi$ and the axion $\chi$), the NS-NS 2-form potential $B$, the R-R 2-form potential $C$ and the R-R 4-form potential $C_{(4)}$. There exists a formulation with manifest SL(2,$\RR$) covariance \cite{Schwarz:1983,Howe:1984}. Here we use the notation of \cite{Cederwall:1997:1,Cederwall:1997:2}. The two 2-forms can be collected in an SL(2,$\RR$) doublet $C_{r}$, where $r$=1,2 corresponds to the NS-NS and R-R 2-forms respectively.  The scalars can be described by a complex doublet ${\cal U}^r$, with $\tau={\cal U}^1/{\cal U}^2=\chi+i\, e^{-\phi}$. The scalar doublet fulfills the SL(2,$\ZZ$)-invariant constraint
\begin{equation}
{\scs \frac{i}{2}}\epsilon_{rs}\,{\cal U}^r\bar{{\cal U}}^s=1
\end{equation}
The 2-form doublet has a 3-form doublet of field strengths $H_{(3)r}=dC_{r}$, which can be combined with the scalar doublet into a complex 3-form
\begin{equation}
{\cal H}_{(3)}={\cal U}^r H_{(3)r}\,,\qquad H_{(3)r}=\epsilon_{rs}\,\mathrm{Im}\big(\,{\cal U}^s\bar{{\cal H}}_{(3)}\big)
\end{equation}
From the scalar doublet we can construct the Mauer-Cartan 1-forms $P$ and $Q$
\begin{equation}
Q={\scs \frac{1}{2}}\,\epsilon_{rs}d{\cal U}^r\bar{{\cal U}}^s\,,\qquad
P={\scs \frac{1}{2}}\,\epsilon_{rs}d{\cal U}^r{\cal U}^s
\end{equation}

The equations of motion can now be written as
\begin{eqnarray}\nonumber
&&D{\ast}P+{\scs \frac{i}{4}}\,{\cal H}_3\wdg\ast{\cal H}_3=0\\\nonumber
&&D{\ast}{\cal H}_3+i\,P\wdg\ast\bar{\cal H}_3-i\,H_5\wdg{\cal H}_3=0\\
&&D{\cal H}_3+i\,\bar{\cal H}_3\wdg P=0\\\nonumber &&dH_5-{\scs
\frac{i}{2}}\,{\cal H}_3\wdg\bar{\cal H}_3=0\\\nonumber &&R_{{\sss
MN}}=2 \bar{P}_{{\sss (M}} P_{{\sss N)}}+{\scs
\frac{1}{4}}\bar{\cal H}_{{\sss (M}}{}^{{\sss RS}}{\cal H}_{{\sss
N)RS}}-{\scs \frac{1}{48}}\,g_{{\sss MN}}\bar{\cal H}_{{\sss
RST}}{\cal H}^{{\sss RST}}+{\scs \frac{1}{96}}\,H_{{\sss
(M}}{}^{{\sss RSTU}}H_{{\sss N)RSTU}} \end{eqnarray} The first two equations
are the equations of motion for $P$ and ${\cal H}_3$,
respectively. The following two are the Bianchi identities for the
3-forms and the 5-form. The last line is the Einstein equations.

%
%
\cleardoublepage
\pagestyle{plain}
\def\href#1#2{#2}
\bibliographystyle{utphysmod}
\bibliography{equivalence0815}

\end{document}